\input{aipcheck}

\documentclass[
    ,final            
  ]
  {aipproc}

\layoutstyle{6x9}

\begin{document}

\title{Causal structure of general relativistic spacetimes}


\author{E. M. Howard}{
  address={Macquarie University, Sydney, Australia}
}

\begin{abstract}
 We present some of the recent results and open questions on the
causality problem in General Relativity. The concept of singularity
is intimately connected with future trapped surface and inner
event horizon formation. We offer a brief overview of the
Hawking-Penrose singularity theorems \cite{Hawking} and discuss a few open
problems concerning the future Cauchy development (domain of dependence) \cite{geroch70}, break-down
criteria and energy conditions for the horizon stability. A key
question is whether causality violating regions, generating a
Cauchy horizon are allowed. 

We raise several questions
concerning the invisibility and stability of closed trapped surfaces
from future null infinity and derive the imprisonment conditions.
We provide an up-to-date perspective of the causal boundaries and
spacelike conformal boundary extensions for time oriented
Lorentzian manifolds and more exotic settings.
\end{abstract}

\maketitle


\section{Introduction}

In this short paper I would like to present a brief overview of some open questions in General
Relativity with dramatic consequences for 
causality theory, aiming to a deeper understanding of the (global) causal structure of the space-time.
Special attention is accorded to the problem of global hyperbolicity, as the most important condition on causality and its relation to time functions. \cite{seifert77}
It is important to note that General Relativity has been infiltrated
with exotic geometries involving closed timelike curves \cite{Thorne}or
other possible causal violations. 
The definition and understanding of the nature of causality
plays a fundamental role in the construction of physical theories.
The number of open issues concerning the causal behaviour of the space-time is fairly large.
Let us recall that the generic conformal invariants that concern causality have been ordered in a "causal ladder" with a few variations regarding the number of properties included in the hierarchy.\cite{minguzzi06c} (See Figure 1)

Roughly speaking, we define causality as the relation between two events corelated in a regular pattern
or between a cause an an effect. All physical theories assume causation as an inherent fundamental assumption.
In relativity, an event can influence another event only if there is a causal (timelike or null) curve connecting the two space-time points.
We may consider 3 different levels of comprehension of the causal structure of the space-time: the first one is an abstract-formal stage, with origins in Special Relativity and assigning a light cone to every single event in space-time. The second stage has a topological nature and considers the local differential behaviour of geodesics on a Lorentzian manifold. The third aspect assumes a cosmological level of our understanding of causality and incorporates classical global problems in General Relativity, as the initial value problem, space-time boundaries or the singularity theorems.   

\section{Remarks}

Is there a current rigorous definition of the causal structure of the space-time?
A generic acceptable definition should contain all the information about
the Lorentzian manifold, causal properties and time orientability, a universal assumption always lying behind any physical theory. The main question about causality and its relation to time is here translated into an issue of assuring a perfect consistency of the causal ladder of space-time.

The most important step in the causal hierarchy is the global hyperbolicity property, located at the top of the ladder. The concept is central to General Relativity and generates several open questions in other context areas: initial value problem, singularity theorems, geodesic inextendability, imprisonment or causal boundaries. Geroch has shown that if a space-time is globally hyperbolic, it admits at least one submanifold that is intersected once by an inextendible timelike geodesic. Briefly, the space-time is globally hyperbolic if it admits a Cauchy hypersurface. Any such a topological surface would be acausal as it is crossed only once by an inextendible timelike curve. In this sense, an achronal causal curve is a lightlike geodesic. One of the other recent results is also the statement that "chronological spacetimes without lightlike lines lines are stable causal" (Minguzzi). The physical meaning of this theorem is the fact that if the space-time admits causality violations, then either the chronology is violated or the space-time is singular. 

The standard definition of global hyperbolicity assumes two separate conditions: 

1. strong causality (no "almost closed causal curves")\cite{Lobo} which was later weakened by Bernal, Sanchez (2000)\cite{bernal06b} by replacing it with causality (no closed causal curves exist), by introducing a new concept of "causal simplicity".

2. compactness of the space-time ("no naked singularities" condition), strictly derived from the Weak Cosmic Censorship Conjecture.   

If we consider causal boundaries over the space-time, other new open questions appear.
It has been proven that a globally hyperbolic space-time always admits a causal boundary containing a timelike subset of causal curves. \cite{seifert71} Again, a causal gradient is implicitly asssumed in order to hold the consistency of the causal theory. 

\begin{figure}
  \includegraphics[height=.3\textheight]{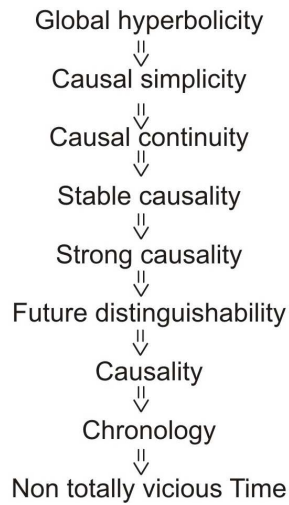}
  \caption{Causal hierarchy of space-time}
\end{figure}

If an unknown factor that defines some sort of continuous increasing function that applies to any future directed causal curve, is applied, the consistency of the causal hierarchy of the spacetime is saved. The implicit or explicit assumption of such a smooth time function containing a pre-ordering timelike gradient \cite{sorkin96} has already been introduced by Hawking in his work "The existence of cosmic time functions".\cite{hawking68} The cosmic time is defined as a global function that increases along every future directed timelike or null curve. The existence of such a function requires causal stability as a fundamental condition (no closed timelike or null curves in any Lorentz metric that is sufficiently near the space-time metric). Hawking proves the equivalence bewtween two fundamental features of the space-time: stable causality and global time.\cite{minguzzi08b} 

Is it necessary to make all these tacit assumptions about a global time function, often associated with a cosmological flow? Can we speak about a global time? Are there any constraints in the local physical laws that would tell us anything about a global time? The concept itself of a global time 
is taken for granted. The time function generates a total pre-ordering gradient on the space-time manifold. It has been proven (Minguzzi) that under physically reasonable conditions, the absence of a global time function would imply a singular space-time. 

\section {Conclusions}

The main problem is that we don't have enough information about the evolution of the space-time manifold from physically reasonable initial conditions and if this evolution would generate naked singularities. There are no clear results that prove the existence of global hyperbolicity. If the space-time is non-globally hyperbolic, the initial Cauchy data can't provide enough information from past time-like infinity to future infinity, to completely determine the current state of the universe. 

If global hyperbolicity doesn't hold, information coming from spatial future null infinity should be taken into account. This concept could work if we modify the notion of causal precedence and disconnect it from the "hidden" timelike gradient self-contained in the causality definition. From a deterministic point of view, the principle of causality states that a physical event described by various variables is fully determined at a given time by a previous event in the causal chain. 

In the new definition, the cause still precedes the effect. However, we separate this expression from a "temporal" point of view. In this case, the cause wouldn't necessarily precede the effect from past to future infinity. Old philosophical definitions of "final causality" could work here. Superluminal causal propagator in scalar quantum field would easily secure this concept. A simple illustration of a non-causal theory is the elliptic Klein-Gordon scalar field, where information can propagate along a closed curve in the space-time, any event along this curve being able to influence itself. Another example could be the scalar field theory with a non-canonical kinetic term (or k-essence), allowing superluminal propagation of information, without threatening causality. 

We basically want to stress here the fundamental role of chronology, orientability and Cauchy problem in the standard definition of causality. We are looking for a generalized definition of causality that wouldn't involve any chronology requirement in its expression (assuming a well posed Cauchy problem). 

The assumption is a satisfactory mechanism that helps to climb the causal ladder, without violating Hawking's chronology protection conjecture \cite{Hawk2}. The "well behaved" Cauchy initial data would indeed contain information from spatial null infinity and it would lead to a "stable" state of the Universe, without contradicting any regular physical assumptions (without violating either chronology or causality). Whether we assume the existence of a global time function or a global hyperbolic space-time, stable causality, as the strongest constraint in the causal ladder would be in this way saved.

\bibliographystyle{aipproc}   

\end{document}